\newcommand{\pT}{$\rm{p_{_{\rm{T}}}}$}
\newcommand{\snn}{$\sqrt{s_{_{\rm{NN}}}}=$}

\documentclass[12pt]{iopart}
\usepackage{graphicx}
\begin{document}

\title[Strangeness and bulk freeze-out properties at RHIC]
{Strangeness and bulk freeze-out properties at RHIC}

\author{A Iordanova (for the STAR Collaboration)}

\address{Department of Physics (M/C 273), University of Illinois at Chicago,
845 W. Taylor Street, Chicago, IL 60607, USA}
\ead{aiorda1@uic.edu}

\begin{abstract}

Identified charged kaon, pion, and proton spectra and ratios from
$\sqrt{\rm s_{_{NN}}}$~=~200 and 62.4~GeV Cu+Cu collisions are studied
with a hydro-motivated blast-wave and a statistical model framework in
order to explore the strangeness production at RHIC and characterize the
bulk freeze-out properties of the created system.  The spectra are
measured at mid-rapidity $(|y|<0.1)$ over the transverse momentum range of
$0.25 < $\pT$ < 1.2$~GeV/$c$ with particle identification derived from the
ionization energy loss in the STAR Time Projection Chamber.

The multi-dimensional systematic study of $\pi^{\pm}$, $K^{\pm}$, 
$p$ and $\bar{p}$ production in Cu+Cu, Au+Au, d+Au and p+p collisions 
is used to discuss the energy, system size and inferred
energy density dependence of freeze-out parameters and strangeness
production.  The new data from Cu+Cu collisions bridge the gap
between the smaller d+Au and larger Au+Au systems, allowing a detailed
study of the onset of strangeness equilibration at RHIC.
\end{abstract}

\vspace{2pc}

\section{Introduction}
Systematic studies of the QCD phase diagram~\cite{cite:QCD_Diagram}
have been enriched by the addition of new RHIC data from Cu+Cu
collisions at \snn~200 and 62.4~GeV.  The STAR experiment has
measured identified particle spectra of $\pi^{\pm}$, K$^{\pm}$,
protons and anti-protons at mid-rapidity $|y|<0.1$ over the transverse
momentum range of $0.25<$\pT$<0.80$~GeV/{\em c} (pions and kaons) and
$0.40<$\pT$<1.20$~GeV/{\em c} for (anti)protons. Comparative analysis
to the previously measured spectra from Au+Au, d+Au and p+p collisions
at RHIC~\cite{200spectra,62spectra} is used to address the energy and
system size effects on freeze-out properties and, particularly, on
strangeness production.

It has been shown that freeze-out parameters in Au+Au collisions at
\snn~200 and 62.4~GeV, have a similar chemical freeze-out temperature at
all centralities and have a decreasing $T_{kin}$ towards more central
 events~\cite{200spectra}.
At the same time the radial flow velocity, $\beta$, increases with centrality. 

The centrality independence of the extracted chemical freeze-out
temperature, indicated that, for different initial conditions,
collisions evolve to the same chemical freeze-out.  For all studied
centralities the values for the chemical freeze-out temperature are
close to the critical temperature, predicted by Lattice QCD calculations,
while changes in $T_{kin}$ and $\beta$ are consistent with higher
energy/pressure in the initial state for more central events. This
suggested that chemical freeze-out coincides with hadronization and
therefore provides a lower limit estimate for a temperature of the prehadronic
state~\cite{Olgaposter}. Most measured bulk properties in Cu+Cu show a smooth
systematic change with the charged hadron multiplicity, and appear to
follow the same systematic trends as the lower-energy
data~\cite{200spectra,62spectra}, bridging the gap between the smaller d+Au 
and the larger Au+Au systems.

\section{STAR Experiment}
The results presented in these proceedings are based on the identification
of charged particles traversing the Time Projection Chamber
(TPC)~\cite{STAR_TPC} in the STAR detector.  Different ionization
energy loss patterns are experienced in the TPC for particles of different
masses, which can be exploited for identification in the low-\pT region.
The momentum measurement is given by the curvature of the particle
trajectories as they pass through a 0.5~T magnetic field.  To determine
the centrality of the collision, the number of charged tracks at
mid-rapidity is used.  The data is presented in six centrality classes
with each bin corresponding to 10\% of the total inelastic cross-section. 

The transverse momentum spectra are obtained from the mean
$\langle dE/dx \rangle$ for each of  $\Delta$\pT~=~50MeV/{\em c}
momentum bins.  For this, projections of the $dE/dx$ for a given
momentum are fit with a four-Gaussian function representing the 
four particle species of a given charge ($\pi$, K, $p$ and $e$). 
The integral of each Gaussian provides the raw yield at a given momenta.  
These raw yields are corrected for detector acceptance, tracking
inefficiency and background contributions.  The same analysis
technique is used for measurements of all different
collision systems and center-of-mass energies~\cite{200spectra}.

\section{Preliminary Results}

\subsection{Particle Spectra}
The $\pi^{\pm}$, $K^{\pm}$ and $p$ and $\bar{p}$ transverse momentum
spectra are measured for two center-of-mass energies in Cu+Cu
collisions,\snn~200 and 62.4~GeV. The particle and anti-particle
spectral shapes are similar for all species at each centrality bin.  At both
collision energies a mass-dependence is observed in the slope of
the particle spectra.

\subsection{Kinetic freeze-out properties}
Within a given centrality bin the particle spectra are fitted
simultaneously by the Blast-wave model~\cite{BlastWaveModel}, which
assumes a radially boosted thermal source.  The hydro-motivated fits
provide information about the radial flow velocity ($\beta$), the
kinetic freeze-out temperature ($T_{kin}$) and the flow profile shape
($n$) at final freeze-out.  The effects from resonance contributions
to the pion spectra shape are reduced by excluding the very low-$p_{T}$
pion data points (below $<$~0.5~GeV/c). 

\begin{figure}[h]
\begin{minipage}{0.5\textwidth}
\includegraphics[width=\textwidth]{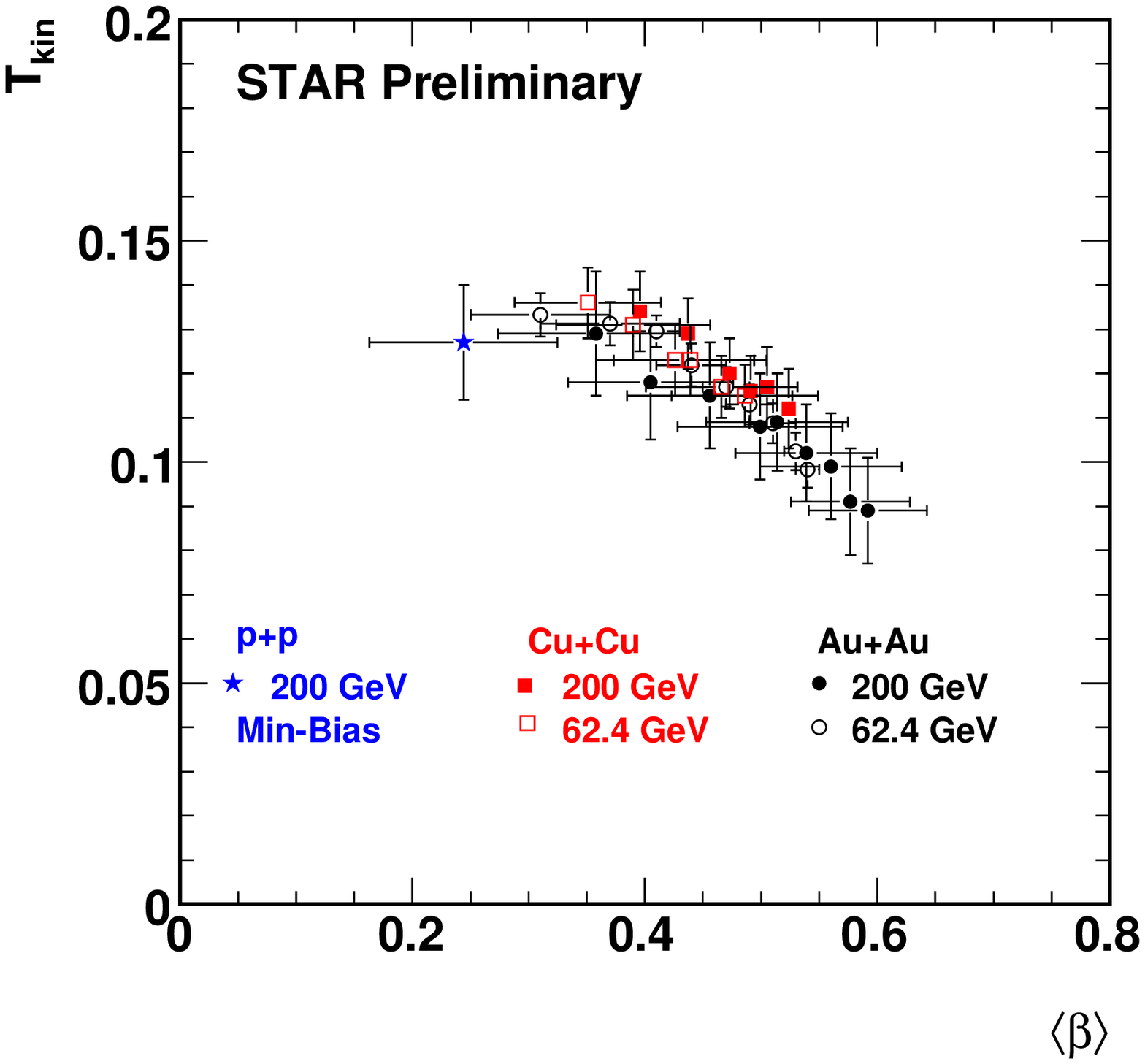}
\end{minipage}
\begin{minipage}{0.5\textwidth}
\includegraphics[width=\textwidth]{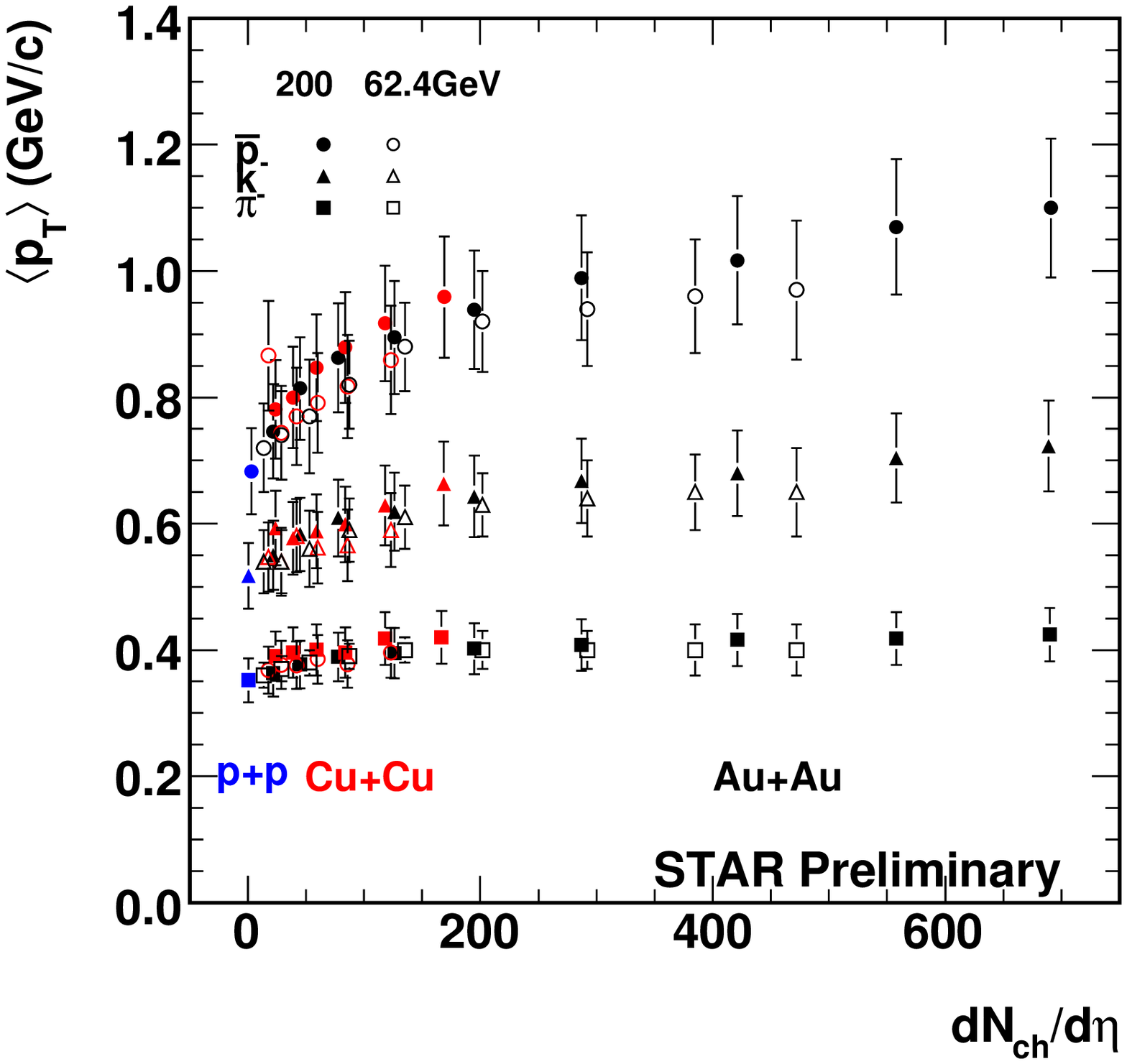}
\end{minipage}
\caption{
\label{fig:KineticFreezeOut}(color online) Left panel: The Kinetic freeze-out
temperature, $T_{kin}$, versus flow velocity, $\beta$ is shown for \snn~200
and 62.4~GeV Au+Au (black) and Cu+Cu (red) collisions.  For comparison, results
for minimum-bias p+p collisions at 200~GeV are also shown. 
Right panel: Integrated $\langle p_{T}\rangle$ for negatively charged 
particles for Cu+Cu (red) and Au+Au (black) 
collisions as a function of $dN_{ch}/d\eta$ for 200 and 62.4~GeV. 
Minimum-bias p+p collisions at 200~GeV are also shown.}
\end{figure}

The particle spectra are well described by a common set of freeze-out
parameters, for all colliding energies. The fit results are shown in
figure~\ref{fig:KineticFreezeOut}, left panel.  
For an equivalent number of charged
particles at mid-rapidity, $dN_{ch}/d\eta$, the $T_{kin}$ and $\beta$
show similar dependences in both Cu+Cu and Au+Au collisions, evolving
smoothly from p+p to central Au+Au.  $T_{kin}$ decreases with centrality
and thus implying freeze-out occurs at lower temperature in more central
collisions (see the right panel on figure~\ref{fig:ChemicalFreezeOut}).  

The particle mean-\pT results are obtained from the measured spectra, 
extrapolated outside the fiducial range  by Blast-wave fits for the
kaons and (anti)proton, and by Bose-Einstein fits for the
pions.  The particle mean-\pT increases with $dN_{ch}/d\eta$
(figure~\ref{fig:KineticFreezeOut}, right panel), 
which is consistent with an increase in radial flow with centrality. 

A model dependent connection between the
number of produced charged particles and the initial gluon density of
the colliding system~\cite{CGC} can be used to interpret that the bulk
properties are most probably determined at the initial stages of the
collision and are driven by the initial energy density.

\subsection{Particle Ratios}
\begin{figure}[h]
\centering
\includegraphics[width=0.5\textwidth]{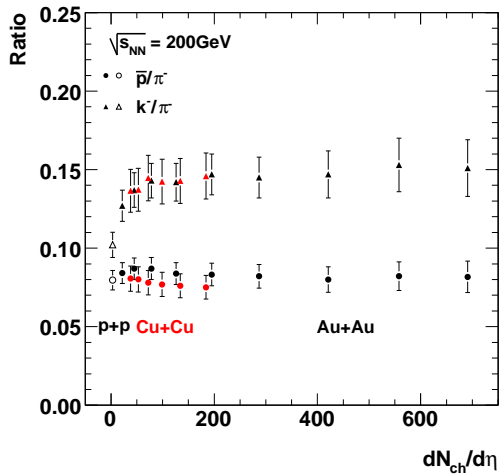}
\caption{\label{fig:ParticleRatios}(color online)
Particle yield ratios at 
200~GeV for Cu+Cu (red) and Au+Au (black) collisions versus 
the number of charged hadron multiplicity density, $dN_{ch}/d\eta$, 
at mid-rapidity. The closed triangles show the $K^{-}/\pi^{-}$ ratios, 
the closed circles are the ratios for $\bar{p}/\pi^{-}$. 
The open symbols are the ratios from min-bias p+p collisions.}
\end{figure}

The ratio of particle yields for negatively charged kaons and pions is
shown in figure~\ref{fig:ParticleRatios} 
for a  center-of-mass energy of
200~GeV.  The $K/\pi$ ratio in the Cu+Cu system follows the same trend
with the charged hadron multiplicity, $dN_{ch}/d\eta$, as previously
found in Au+Au data \cite{200spectra}.  There is no strong evidence for
additional strangeness enhancement in the smaller system as reported at
SPS energies~\cite{SPS_NA49QM02, SPS_NA49PRL}, despite the observed
increase in the integrated particle spectra yields with respect to p+p
data for a given value of $N_{part}$~\cite{ATimmins}.
The baryon to meson ratios, also shown on the 
figure~\ref{fig:ParticleRatios}, are found to be the same for Au+Au and
Cu+Cu systems.  The lack of a strong centrality dependence over the
covered range points to similar freeze-out conditions for the studied
collisions.

\subsection{Chemical freeze-out properties}

The particle yield ratios are further analyzed within the framework of the
statistical model~\cite{StatModel}.  The model describes the chemical
freeze-out properties of the colliding system by the chemical freeze-out
temperature ($T_{ch}$), the baryon and strangeness chemical potentials
($\mu_{B}$, $\mu_{S}$) and the strangeness suppression factor ($\gamma_{S}$).  
The parameters are obtained using only 
$\pi^{\pm}$, $K^{\pm}$, $p(\bar{p})$ measurements.

Within the systematic errors on the fit
parameters the strangeness suppression factor $\gamma_{S}$ in Cu+Cu,  
is consistent with the results for the
Au+Au data~\cite{200spectra}.  
This parameter shows a similar dependence with  $dN_{ch}/d\eta$,
 as in the Au+Au system.
The values of $\gamma_{S}$ approaching unity for the central collisions 
implies that the produced
strangeness is close to approximate equilibrium.

\begin{figure}[h]
\begin{minipage}{0.5\textwidth}
\includegraphics[width=\textwidth]{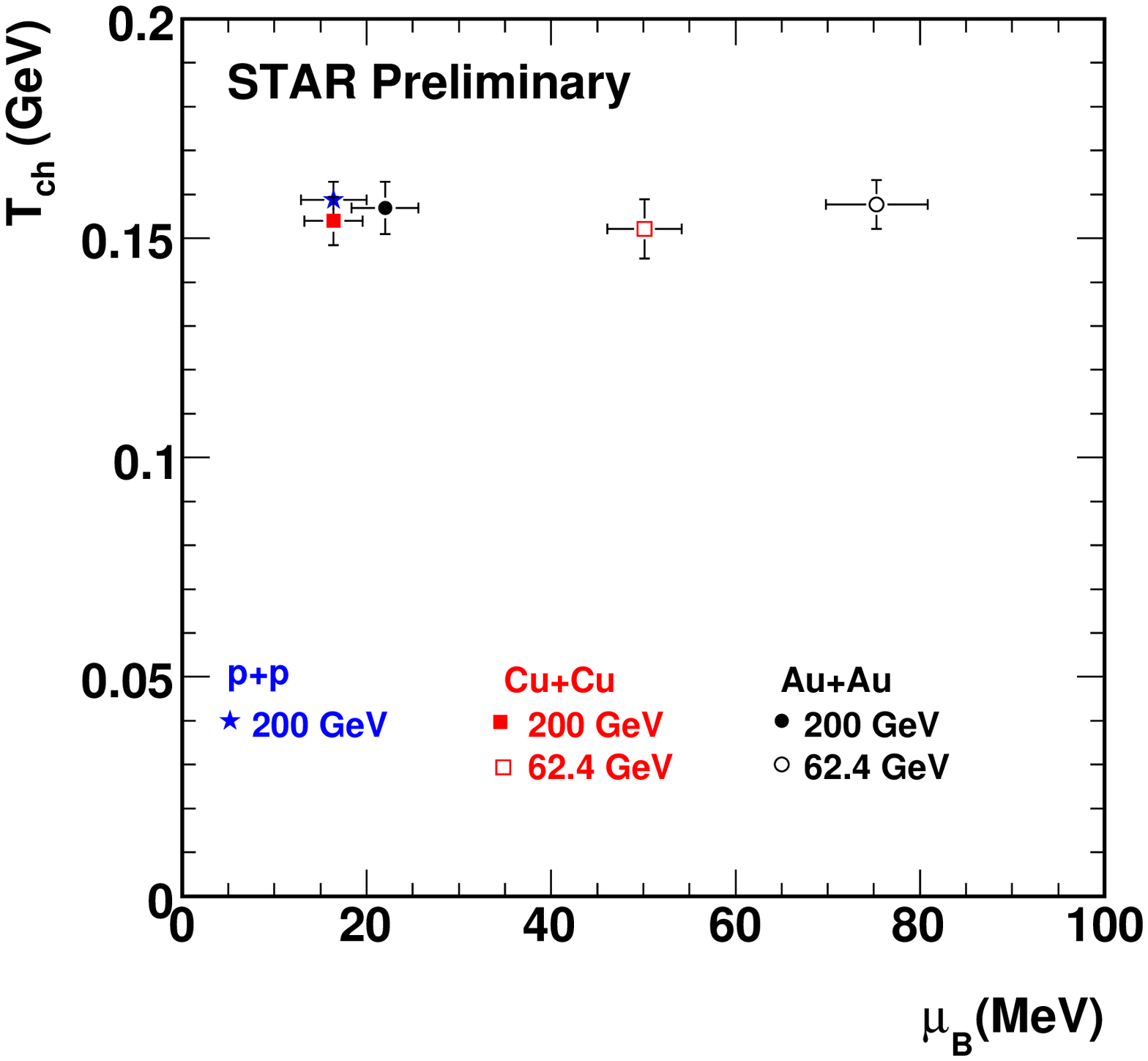}
\end{minipage}
\begin{minipage}{0.5\textwidth}
\includegraphics[width=\textwidth]{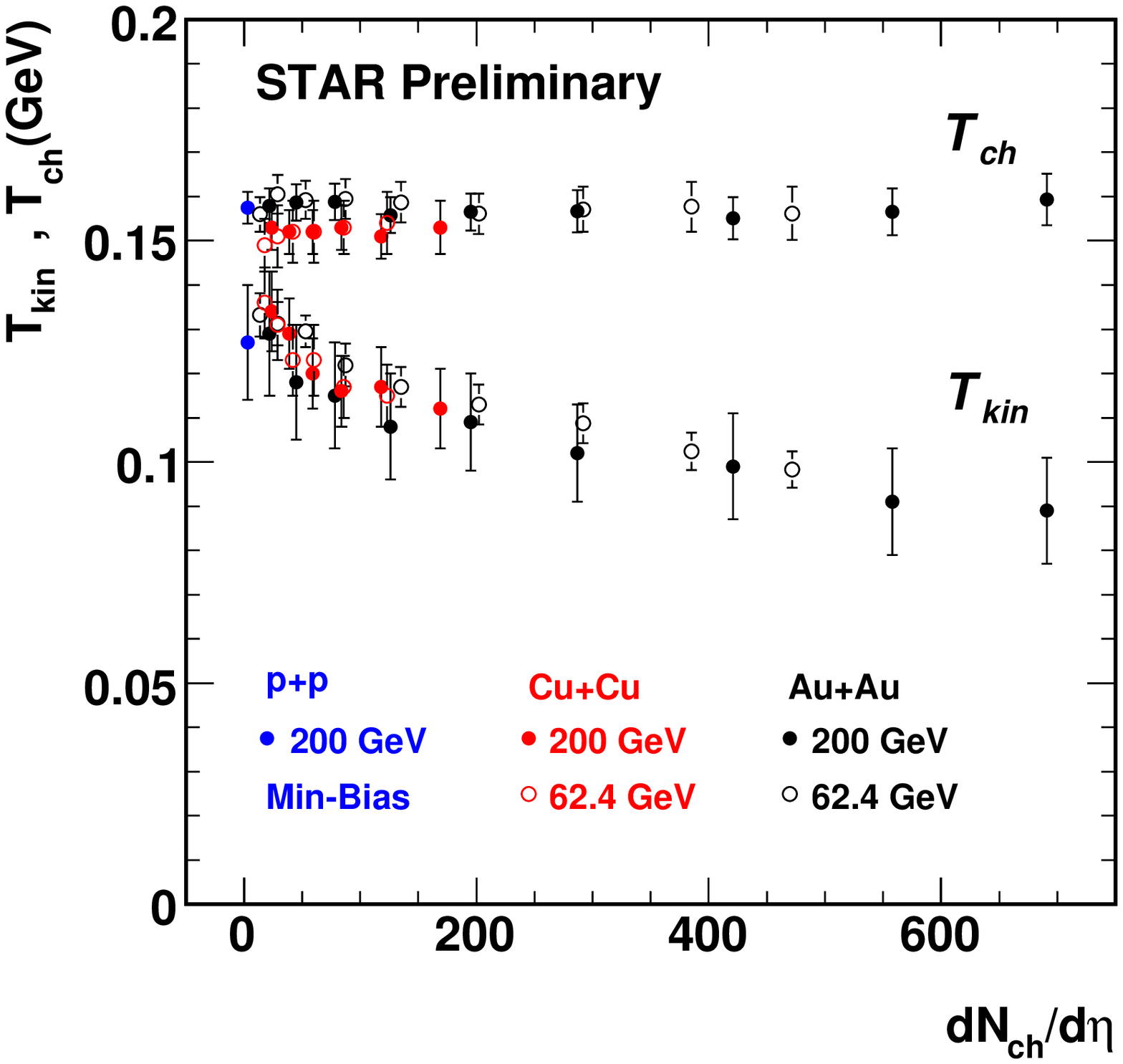}
\end{minipage}
\caption{\label{fig:ChemicalFreezeOut}(color online) 
Left panel: The Chemical freeze-out 
temperature, $T_{ch}$, versus the  baryon chemical potential, $\mu_{_{B}}$, for
central Au+Au (0-5\% - black) and Cu+Cu collisions (0-10\% - red). 
Minimum-bias p+p data at 200~GeV are also shown (blue).
Right panel: $T_{ch}$($T_{kin}$) versus charged hadron 
multiplicity  at \snn~62.4 (open symbols) and 200~GeV (closed symbols) for 
Cu+Cu (red) and Au+Au (black) collisions. For comparison, results for 
minimum-bias p+p collisions at 200 GeV are also shown (blue).}
\end{figure}

The chemical freeze-out temperature, $T_{ch}$, as a function of baryon-chemical
potential, $\mu_{B}$, for different systems is shown in the left panel of
figure~\ref{fig:ChemicalFreezeOut}.  
For all systems and center-of-mass energies
$T_{ch}$ appears to be universal.  The value of the baryon chemical potential
reflects the decrease in baryon density from $\sqrt{s_{NN}}=$~62.4 to 200~GeV.
At an equivalent center-of-mass energy $\mu_{B}$ is higher for the larger
system.  The constant value of $T_{ch}$ implies that collisions with different
net-baryon densities evolve to the same chemical freeze-out and points to 
a universal hadronization of the system.

\section{Summary}
The STAR collaboration has presented measurements of identified charged
hadron spectra in Cu+Cu collisions for two center-of-mass energies, 200
and 62.4~GeV.  These new results of $\pi^{\pm}$, $K^{\pm}$, $p(\bar{p})$
have further enriched the variety of low-$p_{T}$ spectra at RHIC.
The data have been studied within the statistical and Blast-wave model
frameworks in order to characterize the properties of the final hadronic
state of the colliding system as a function of system size, collision
energy, centrality and the inferred energy density. 

This multi-dimensional systematic study reveals remarkable similarities
between the studied systems.  The obtained particle ratios, mean-\pT~and
the freeze-out parameters, including the strangeness saturation factor
$\gamma_{s}$, are found to be intrinsically related for all
 collision systems and center-of-mass energies.  A smooth evolution
with $N_{ch}$ and similar properties at the same number of produced
charged hadrons are observed.  A model dependent connection between the
number of produced charged particles and the initial gluon density of
the colliding system~\cite{CGC} can be used to interpret that the bulk
properties are most probably determined at the initial stages of the
collision and are driven by the initial energy density.

\section*{Acknowledgments}
We thank the RHIC Operations Group and RCF at BNL, and the
NERSC Center at LBNL for their support. This work was supported
in part by the Offices of NP and HEP within the U.S. DOE Office 
of Science; the U.S. NSF; the BMBF of Germany; CNRS/IN2P3, RA, RPL, and
EMN of France; EPSRC of the United Kingdom; FAPESP of Brazil;
the Russian Ministry of Sci. and Tech.; the Ministry of
Education and the NNSFC of China; IRP and GA of the Czech Republic,
FOM of the Netherlands, DAE, DST, and CSIR of the Government
of India; Swiss NSF; the Polish State Committee for Scientific 
Research; Slovak Research and Development Agency, and the 
Korea Sci. \& Eng. Foundation.


\section*{References}
\begin{thebibliography}{10}

\bibitem{cite:QCD_Diagram} F.Karsch, {\it J.Phys.Conf.Ser.}  {\bf 46} (2006) 122.

\bibitem{200spectra} J.Adams et al., {\it Phys. Rev. Lett.} {\bf 92} (2004) 112301.

\bibitem{Olgaposter} O.Barannikova et al., {\it arXiv:} nucl-ex/0403014. 

\bibitem{62spectra} L.Molnar et al., {\it arXiv:} nucl-ex/0507027.


\bibitem{STAR_TPC} M. Anderson et al., {\it Nucl. Instrum. Meth.} {\bf A499} (2003) 659.

\bibitem{BlastWaveModel} E.Schnedermann, J.Sollfrank and U. Heinz, {\it Phys. Rev.} {\bf C48} (1993) 2462.

\bibitem{CGC} L.McLerran, {\it Acta Phys. Polon.} {\bf B34 } (2003) 3029; D.Kharzeev and E.Levin, {\it Phys. Lett.} {\bf B523 } (2001) 79; D.Kharzeev, E.Levin, L.McLerran {\it Phys. Lett.} {\bf B561 } (2003) 93.


\bibitem{SPS_NA49QM02} S.V.Afanasiev et al., {\it Nucl. Phys.} {\bf A715 } (2003) 474, {\it arXiv:} nucl-ex/0209018v1.

\bibitem{SPS_NA49PRL} C. Alt et al. {\it Phys. Rev. Lett.} {\bf 94} (2005) 052301, {\it arXiv:} nucl-ex/0406031v2.


\bibitem{ATimmins} A. Timmins et al., {\it these proceedings.}

\bibitem{StatModel} P.Braun-Munzinger, I.Heppe and J.Stachel, {\it Phys. Lett.} {\bf B465} (1999) 15.


\end{thebibliography}
\end{document}